\documentclass{moriond}

\def\be{\begin{equation}}
\def\ee{\end{equation}}
\def\bea{\begin{eqnarray}}
\def\eea{\end{eqnarray}}

\newcommand{\Photo}{\includegraphics[height=33mm]{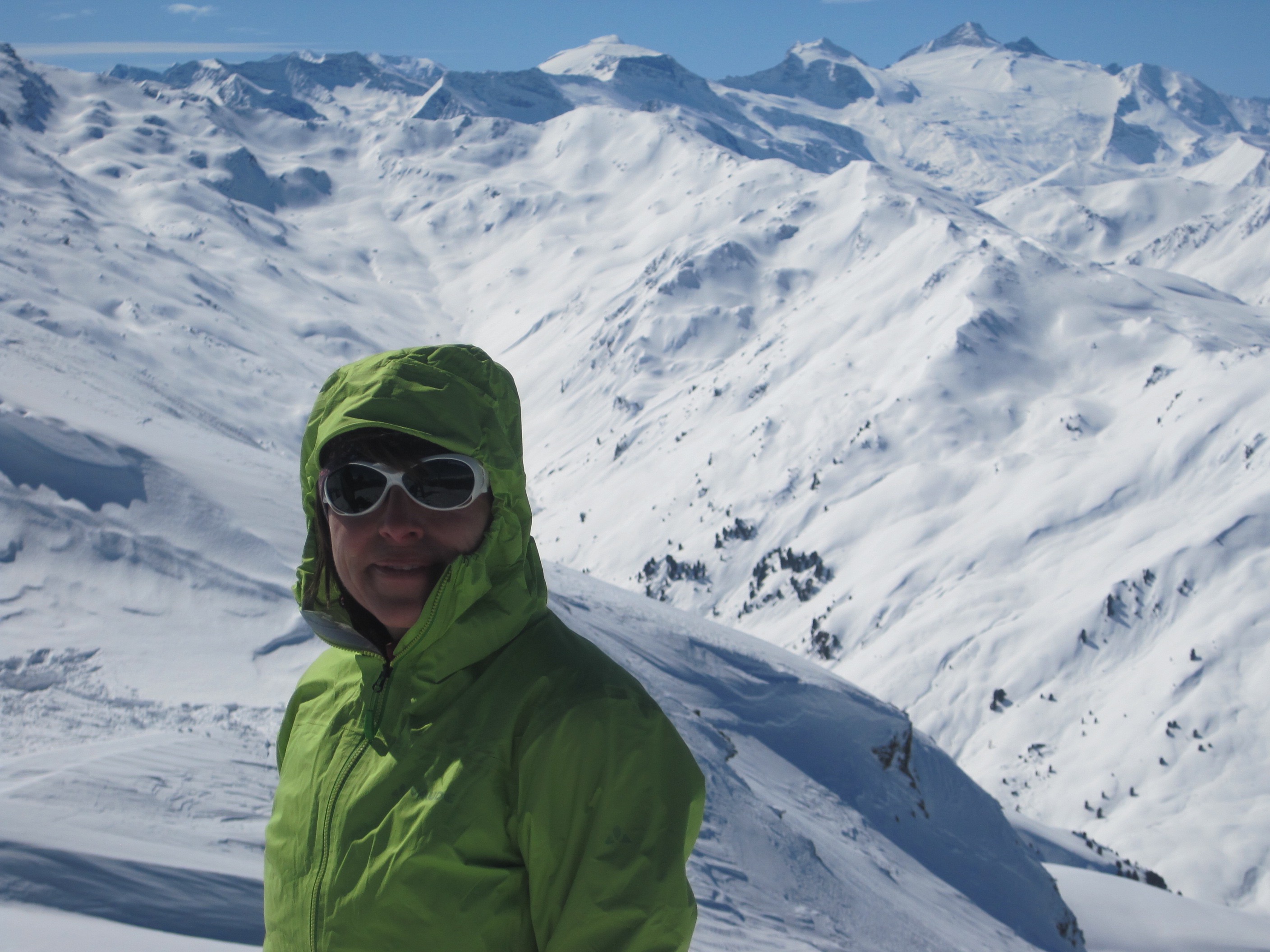}}

\begin{document}
\vspace*{4cm}
\title{QCD AND HIGH ENERGY INTERACTIONS: \\MORIOND 2018 THEORY SUMMARY}

\author{Gudrun Heinrich}

\address{Max Planck Institute for Physics, F{\"o}hringer Ring 6, 80805
Munich, Germany}

\maketitle
\abstracts{
The highlights of the theory developments presented at the Rencontres de Moriond 2018 on QCD and High
Energy Interactions are summarised and put into perspective.}

\section{Introduction}

With the wealth of new experimental results presented at Moriond 2018 and still to
come, theorists have to keep up with the increasing experimental
precision, offer interpretations of the data, 
come up with new ideas how to probe the Standard Model and
think ahead what could be beyond and how it could be tested.
All these points have been addressed in may exciting talks, reflecting
the current situation in particle physics from the QCD side, whose
understanding in all aspects is of great importance, both in its
own right and also in order to tell apart New Physics from QCD effects.

\section{Higgs Physics and precision calculations}


As the excitement about the Higgs boson discovery is ebbing away, we
should not forget that we are just at the beginning of our exploration
of the Higgs sector. The fact that the Higgs boson so far looks pretty
Standard-Model-like means that higher order corrections in the Higgs
sector are extremely important in order to establish that small
deviations form the Standard Model (SM) predictions are indeed signs of New
Physics. Therefore Higgs physics and precision calculations are
closely related.
As Keith Ellis~\cite{Ellis} put it, new results in the Higgs sector are
``guaranteed deliverables'', and therefore it should be our primary
goal to scrutinise the Higgs sector, in particular get a handle on the
Higgs couplings to light SM particles, the total and partial widths, invisible
decays and the trilinear Higgs coupling. He also presented a
comparison of various future collider options and put it into
perspective for the upcoming European Particle Physics Strategy document.

\begin{figure}[htb]
\begin{minipage}{0.6\linewidth}
\centerline{\includegraphics[width=0.7\linewidth]{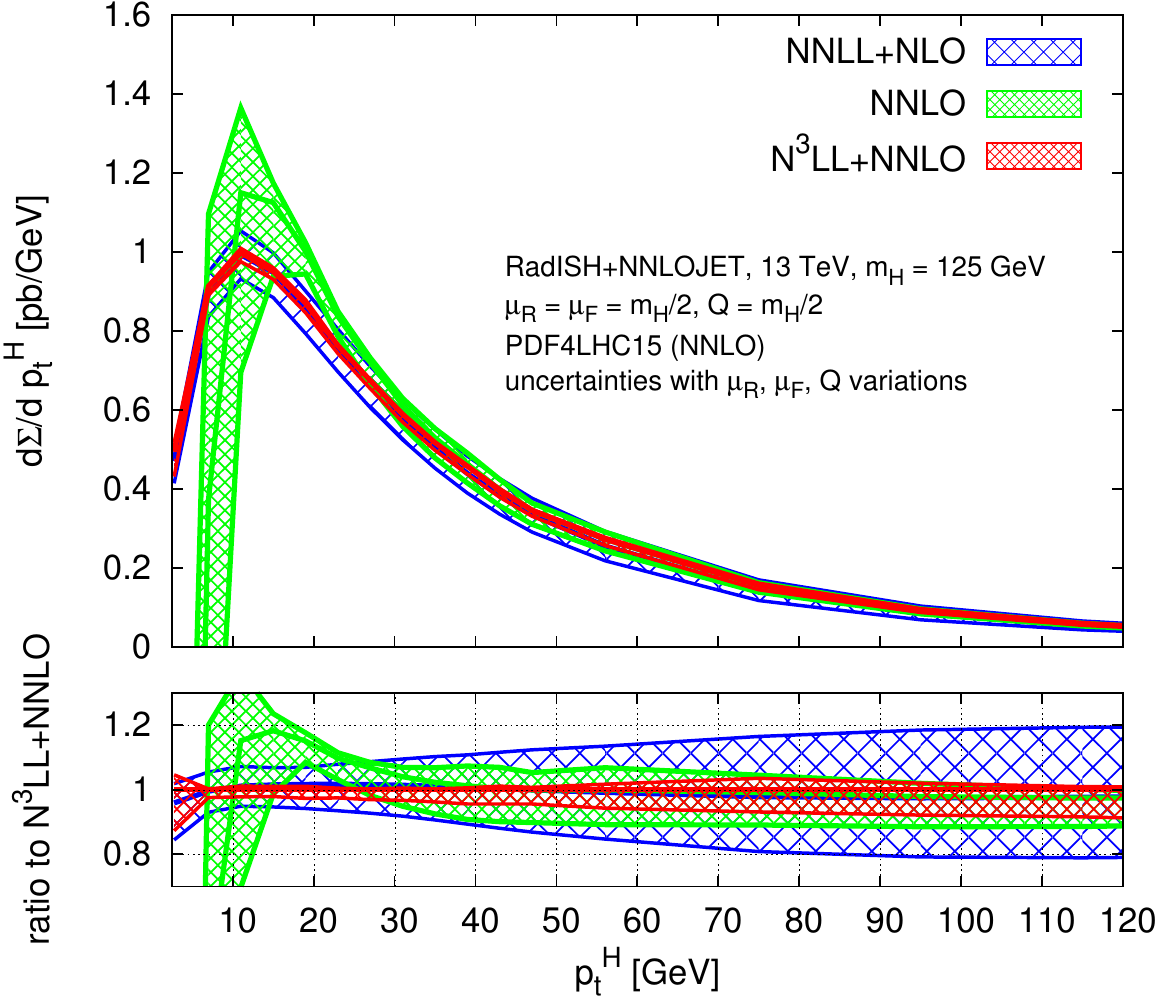}}
\end{minipage}
\hspace*{-2.5cm}
\begin{minipage}{0.6\linewidth}
\centerline{\includegraphics[width=0.7\linewidth]{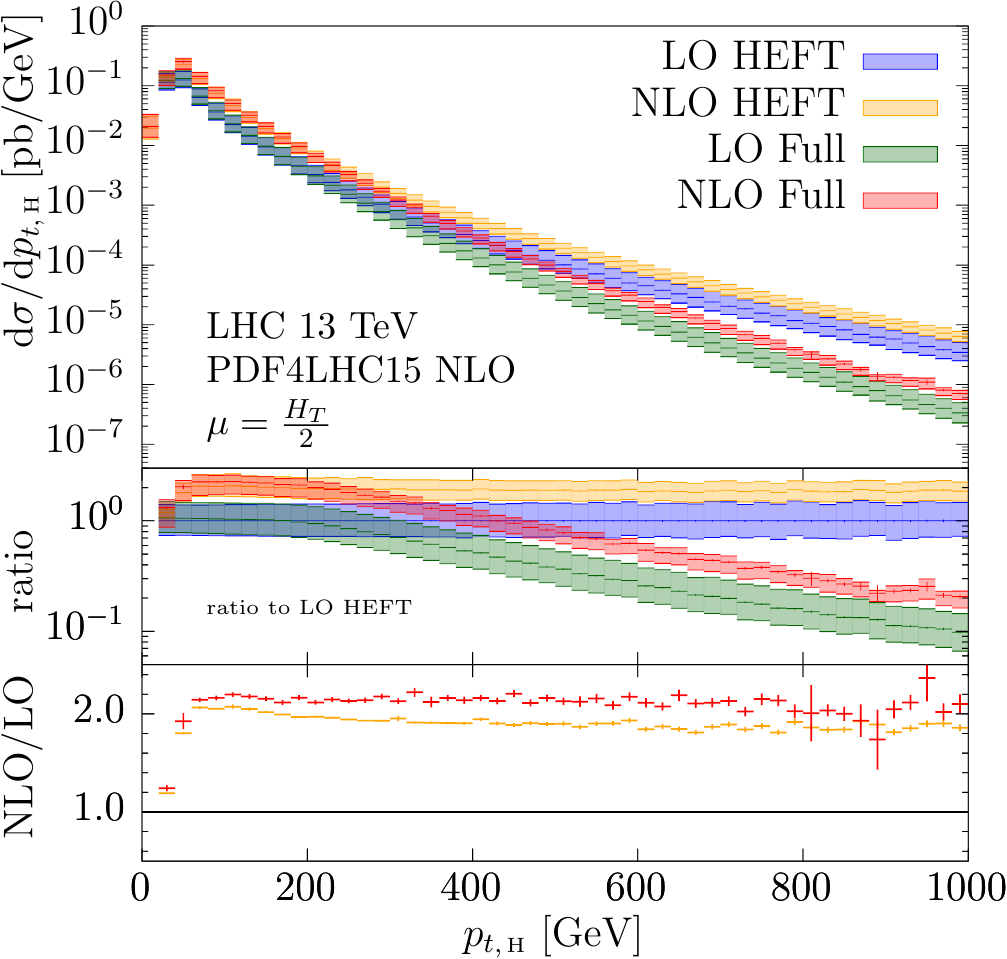}}
\end{minipage}
\caption[]{Left:  comparison of the transverse momentum distribution for Higgs boson production at NNLO and
N$^3$LL+NNLO, NNLL+NLO, and NNLO for a central scale choice of $\mu_R = \mu_F = m_H/2$.
Right: Higgs boson transverse momentum spectrum at LO
and NLO  QCD in the $m_t\to\infty$ limit (HEFT) and with full top-quark mass dependence.
The upper panel shows the differential cross sections,
the middle panel the ratio to
the LO HEFT prediction, the lower panel  the
differential K-factors for both the HEFT (orange) and the full theory (red).}
\label{fig:radish}
\end{figure}

An update on available predictions and their uncertainties for Higgs
boson production and decay was given by Michael Spira~\cite{Spira}.
While for inclusive Higgs boson production the theoretical
uncertainties nowadays are rather well under
control~\cite{deFlorian:2016spz,Dulat:2017prg,Mistlberger:2018etf}, 
this was less the case for the Higgs boson transverse momentum. 
Important progress on this subjet was presented at Moriond 2018. 
Luca Rottoli reported on results~\cite{Rottoli,Bizon:2017rah} based on momentum space resummation
at N$^3$LL, matched to NNLO in the heavy top limit, for the Higgs
boson $p_T$ spectrum, see Fig.~\ref{fig:radish}(a).
The same procedure also has been used to produce predictions at this
level of precision for the Drell-Yan process, and the method allows to
resum entire classes of observables.
Results for the Higgs boson transverse momentum spectrum at  N$^3$LL,
matched to NNLO also have been presented very recently in
Ref.~\cite{Chen:2018pzu}.
However, the heavy top limit is not a good approximation in the 
$p_T$-range where the energy is sufficient to resolve the top quark
loops. Therefore the prediction of the Higgs boson $p_T$-spectrum at
NLO with full top quark mass dependence was much in demand, 
and has been presented by Matthias Kerner~\cite{Kerner}, showing that
the full result differs from the NLO result in the heavy top
approximation by about 9\% at total cross section
level~\cite{Jones:2018hbb}. The full top quark mass dependence
significantly alters the tail of the $p_{T,H}$-distribution, as can be
seen from Fig.~\ref{fig:radish}(b).
Note that the top-bottom interference effects for Higgs+jet at NLO are
also available~\cite{Lindert:2017pky}.

\hspace*{-2cm}

Precision calculations within models suggesting an extended Higgs
sector, for example the Two-Higgs Doublet Model, are important for BSM
searches. Stefan Dittmaier reported on a calculation of NLO
electroweak and QCD corrections 
to the decay $h \to WW/ZZ \to 4$\,fermions of the light CP-even Higgs
boson within various types of Two-Higgs-Doublet
Models, available in {\tt  PropHecy4f}~\cite{Dittmaier,Altenkamp:2017kxk},
comparing also various renormalisation schemes.

Certainly, model independent approaches to the search for non-SM
phenomena in the Higgs sector are also very important, and many
creative ideas, also including deep learning algorithms, are emerging
rapidly these days. 
For example, optimisations of measurements in the Higgs sector
based on information geometry, was presented by Felix
Kling~\cite{Kling,Brehmer:2016nyr}.

\section{Perturbative QCD at work}

In the discussion session with the title ``Where is New Physics?'' the
question was raised how well ``QCD backgrounds'' are under control. 
The answer is: better and better! At Moriond 2018, the record of loops+legs was held by
Ben Ruijl (5 loops, 2 legs) and Ben Page (5 legs, 2 loops). Ben Page
presented results for planar 2-loop 5-gluon
amplitudes~\cite{Page,Abreu:2017hqn}, obtained by numerical unitarity
and finite field reconstruction,
thus using a method which may replace standard two-loop reduction methods
at some point.

Ben Ruijl~\cite{Ruijl} presented a generalised $R^\star$
operation~\cite{Herzog:2017bjx} to extract pole parts from multi-loop Feynman
graphs with numerators, which has been worked out in order to
calculate the 5-loop $\beta$-function~\cite{Herzog:2017ohr}.
He also showed results for Higgs decays to gluons at
N$^{4}$LO~\cite{Herzog:2017dtz}, which reduce the scale uncertainties
by almost a factor of 4 compared to N$^{3}$LO, such that the
uncertainties due to the truncation of the perturbative series now play
a sub-leading role in $\Gamma_{H\to gg}$ compared to other effects
(e.g. top mass effects, $\alpha_s$).

Fabrizio Caola~\cite{Caola} reported about an efficient method to handle infrared
divergent real radiation at NNLO, called ``nested soft-collinear
subtraction''~\cite{Caola:2017dug}, which has been used to calculate
$pp\to WH, H\to b\bar{b}$ at NNLO~\cite{Caola:2017xuq}.
Large corrections have been found at NNLO in regions which are not
populated by LO, as well as interesting effects related to the fact that the
$b$-quark is treated as massless.

Rhorry Gauld presented a study of the angular coefficients which
parametrise the angular dependence of the decay to leptons in $Z$-boson production at
NNLO~\cite{Gauld,Gauld:2017tww}, based on a calculation of $Z+X$ in
the {\sc NNLOJet} framework~\cite{Ridder:2015dxa}. These coefficients
have been measured and a tension with previous predictions has been
observed for $A_0-A_2$. It turned out that NNLO is necessary to
describe the data.

David d'Enterria revisited the forward-backward asymmetry of
$b$-quarks in $e^+e^-\to Z\,(\to b\bar{b})$,
asking the question whether the large ($2.8\sigma$) discrepancy between data and
theory predictions in $A_{FB}^{0,b}$, determined at LEP times, persists if the QCD
uncertainties are re-assessed with modern simulation tools~\cite{dEnterria:2018jsx}.
Interestingly it turns out that the QCD uncertainties
 are overall slightly smaller but still consistent with the original ones.

\section{PDFs, non-perturbative QCD and Spectroscopy}

Cranking up the orders in perturbative QCD is useful only if the gain
in precision is not spoiled by non-perturbative effects. 
Major contributions to this subject concern PDF determinations  and a
better understanding of hadron physics.

\subsection{New developments related to PDFs}

German Sborlini showed us that QED effects can compete with NNLO QCD
effects and therefore it is important to take a photon content of
the proton into account, as  has been pointed out in~\cite{Manohar:2016nzj}.
He presented a calculation of the mixed QED-QCD splitting functions~\cite{deFlorian:2016gvk}
and also showed a calculation of NLO QED corrections to diphoton
production, pointing out that jet vetos can lead to enhanced QED effects~\cite{Sborlini:2018fhr}.

R.~Zlebcik presented a new method to calculate transverse momentum
dependent parton distribution functions and its implementation into
{\tt xFitter}~\cite{xfitter}, together with applications to the
$p_{T}$ distribution of the $Z$-boson and
dijet decorrelations~\cite{Zlebcik}. The version {\tt xFitter-2.0.0}
and its various functionalities has been presented by
F.~Giuli~\cite{Giuli:2018rmm}, focusing in particular on the inclusion of
{small-$x$} resummation. 
J.~Fiaschi pointed out that the forward-backward asymmetry from
Drell-Yan production could also be useful to be included in PDF fits~\cite{Fiaschi:2018buk}.
Promising results have been shown from the Lattice community by
S.~Zafeiropoulos~\cite{Orginos:2017kos}. A new approach based on ratios
of matrix elements leads to quark PDFs from first principles which
are competitive with PDFs from global fits.

\subsection{Non-perturbative QCD and hadron physics}

News about modelling quantum effects in hadronisation, in particular on
a model which provides an explanation for the emergence of
Bose-Einstein-like correlations without additional free parameters, have been reported
by S.~Todorova-Nova~\cite{Todorova,Todorova-Nova:2018uat}.
B.~Kerbikov proposed a dynamic model for sound absorption and bulk viscosity
near the critical temperature~\cite{Kerbikov:2018wdl,Kerbikov:2016lqr}.


Chris Quigg pointed out that stable heavy tetra-quark mesons of type
$Q_iQ_j\bar{q}_k\bar{q}_l$ must exist in the limit of very heavy
quarks $Q$. 
He predicts that double-beauty states will be stable against strong
decays, while the double-charm states 
and mixed beauty-charm states will dissociate into pairs of heavy-light mesons.
Observation of such states would establish the existence of
tetra-quarks, and comparison with
theoretical predictions for their production rate and lifetime would
teach us about the role of heavy colour-antitriplet di-quarks as hadron constituents~\cite{Quigg:2018eza}.

Shi-Yuan Li also pursued the deeper understanding of multi-quark states  and
emphasised the importance of studying hadron correlations in this context~\cite{Li}.

An explanation why the lifetimes of five $\Omega_c$ excited states --
recently found by LHCb and confirmed by Belle --
are so small, has been offered by
M.~Praszalowicz~\cite{Praszalowicz:2018azt}, based on an extension of the chiral
quark-soliton model, where the two narrowest states are interpreted as
penta-quarks belonging to the $\overline{\bf 15}$ representation of $SU(3)$.

Cai-Dian Lu offered calculations of branching fractions for several
doubly charmed baryon states~\cite{Li:2018epz}, and in fact the $\Sigma_{cc}^{++}$ state has
been discovered by LHCb in the decay channel
$\Lambda_c^+K^-\pi^+\pi^+$, 
which has been calculated to have the largest branching fraction.

\section{Flavour Physics}

The flavour sessions started with a talk by Gudrun Hiller\cite{Hiller:2018ijj} stating the
experimental facts and their possible theory implications: 
there are hints in semi-leptonic B-meson decays ($b \to s ll$ transitions)
pointing towards a violation of lepton-universality. The ratios
\begin{equation}
 R_{H} =\frac{ \int_{\rm q^2_{\rm min}}^{q^2_{\rm max}} d q^2 \, d{\cal{B}}/dq^2 (\bar B \to \bar H \mu \mu) }{ \int_{\rm q^2_{\rm min}}^{q^2_{\rm max}} d q^2  \, d {\cal{B}}/dq^2  (\bar B \to \bar H e e) }  
\end{equation}
for $H=K$ and $K^*$ have been measured to deviate from unity at the
$\sim 2.6\,\sigma$ level~\cite{Crocombe,Albrecht}, while the radiative corrections do not exceed
the percent level. 
The dimension six operators which can be responsible for a violation
of lepton-universality can be clearly identified~\cite{Hiller:2014ula,Hiller:2003js}.
The measured ratios  $R_{D}$ and  $R_{D^*}$ ($b \to c l\nu$
transitions) also seem to indicate signs of lepton-non-universality (LNU). 
Rather minimal extensions of the Standard Model to account for LNU would
be $U(1)$ extensions ($Z^\prime$-models) with gauged lepton
flavour~\cite{Altmannshofer:2014cfa}  or  leptoquarks, see
Section~\ref{subsec:flavour_anomalies}.
Collider searches for such states will certainly give important
information complementary to the data from LHCb and Belle II.

Robert Fleischer~\cite{Fleischer:2018qcs} reported on a theoretical framework to study leptonic
decays $B_q^0\to l^+l^- (q=s,d)$, which belong to the cleanest
rare $B$ decays,  and therefore offer an outstanding opportunity to explore the 
flavour sector. So far, only $B_s^0\to \mu^+\mu^- $ has been observed,
and agrees with the Standard Model expectation. On the other hand, another
promising decay, $B_s^0\to e^+e^- $, has received little attention so far
because of its  helicity suppressed Standard Model branching ratio,
which however may be
hugely enhanced through new (pseudo)-scalar contributions which lift
the helicity suppression. He also pointed out that new sources of CP
violation may enter the game, and presented observables which are well
suited to explore in this direction, while not giving up flavour universality~\cite{Fleischer:2018qcs,Fleischer:2017yox}.
Utilising $B\to \pi K$ decays as a probe of new physics, in particular
with regards to CP asymmetries, is another interesting subject. Ruben
Jaarsma has presented a new state-of-the-art analysis, including also
effects from electroweak penguin diagrams, which confirms the
tension with current data~\cite{Fleischer:2018lcs,Fleischer:2018bld,Fleischer:2017vrb}.

Roman Zwicky explained that the contamination of right-handed currents
in $B\to V\gamma$ (or $B\to V\,l\bar{l}$)
decays due to long-distance effects can be controlled by considering in addition the
corresponding decay $B\to A\gamma\,(l\bar{l})$,  where $V/A$ are vector/axial vector mesons,
exploiting the opposite relative sign of left- versus right-handed amplitudes~\cite{Zwicky,Gratrex:2018gmm}.

Giancarlo D'Ambrosio gave us a broad overview on recent developments
in Kaon physics~\cite{Ambrosio}, also listing models which address
the $\epsilon^\prime/\epsilon$ anomaly and discussing the interplay
with B-anomalies. 
More details will be provided in the following section.

\section{Beyond the Standard Model}
\label{subsec:flavour_anomalies}

The contributions about physics beyond the Standard Model can roughly
be divided into two categories: the ones which address specifically the
flavour anomalies, and the ones which don't. 

In the first category (see also~\cite{Crivellin:2018gzw} for a small review) is a class of models presented by
Andreas Crivellin~\cite{Crivellin,Blanke:2018sro}, which is of
Pati-Salam type, i.e. based on a gauge group where lepton number is
the fourth colour. Implementing this
gauge symmetry  in a 5D Randall-
Sundrum  background, the mass scales of the
Kaluza-Klein  resonances, in particular the vector leptoquarks, can
be in the few TeV range. The model naturally accommodates the
$R_{K,K^*}$ anomalies, and in a non-minimal version also can offer an explanation of the tensions in the anomalous
magnetic moment of the muon.

Abhishek Iyer also offered an explanation for the anomalies in $B$-
and rare $K$-decays, in the context of 
custodial Randall-Sundrum models~\cite{Iyer:2018zvf,DAmbrosio:2017wis}. 
Two solutions are possible within such models, one where both muons
and electrons play a role in lepton non-universality, and one where
primarily muons play a role. More data on rare Kaon decays could serve
to distinguish the two possibilities~\cite{Ambrosio}.

\begin{figure}
\begin{minipage}{0.6\linewidth}
\centerline{\includegraphics[width=0.7\linewidth]{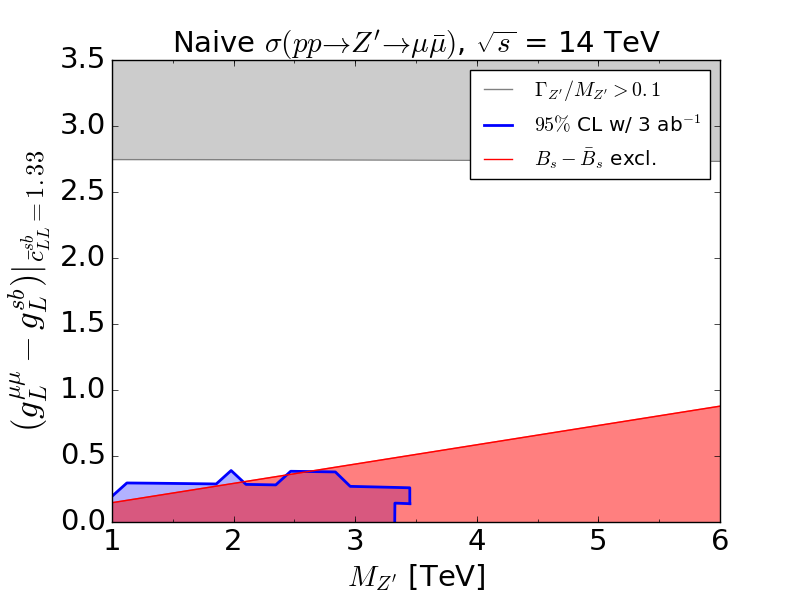}}
\end{minipage}
\hspace*{-2cm}
\begin{minipage}{0.6\linewidth}
\centerline{\includegraphics[width=0.7\linewidth]{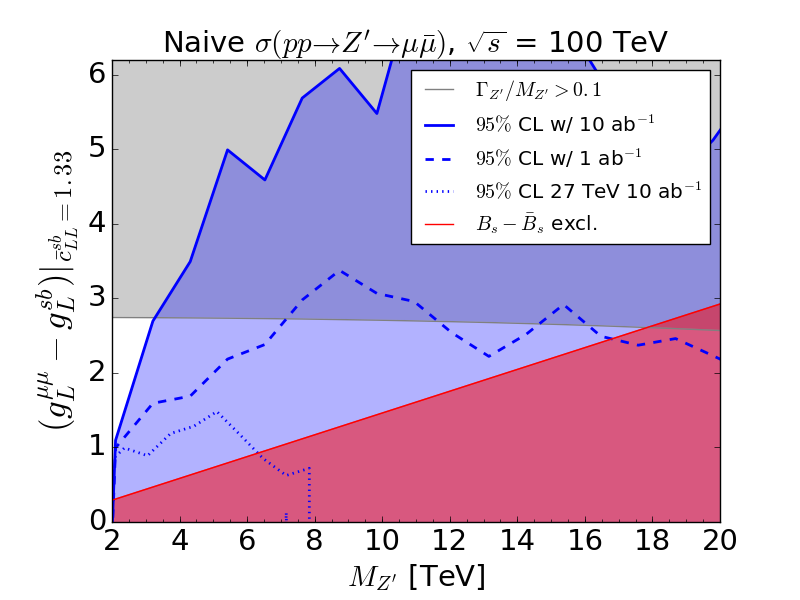}}
\end{minipage}
\caption{Coverage of the parameter space for a minimal
  $Z^\prime$-model that could explain the anomaly in $B\to K^{(*)}\mu^+\mu^-$. 
Left: HL-LHC, right: FCC at 100 TeV.}\label{fig:discovery_potential}
\end{figure}

What is next if (some of) the indirect signs of New Physics in the flavour
sector turn out to be firmly established?
Tevong You addressed the interesting question whether we can reach the
scale of  New Physics which may be behind the flavour anomalies at
future colliders~\cite{You:2018yno,Allanach:2017bta}.
Focusing on rather minimal $Z^\prime$ and leptoquark models, 
the conclusion is that for narrow width $Z^\prime$ models there is a good but not
complete sensitivity at the HE-LHC, whereas  FCC-hh would almost
guarantee a discovery, see Fig.~\ref{fig:discovery_potential}.
If leptoquarks are responsible, the conclusion depends critically on
the leptoquark masses, but  for masses below 4.5 TeV a discovery at
HE-LHC would be very likely.


Extended Higgs sectors can offer solutions to open questions like the
origin of dark matter, baryogenesis or unexplained hierarchies.
Howie Haber gave a classification of extended Higgs sectors within the
framework of two-Higgs doublet models
(THDMs)~\cite{Haber:2018ltt,Draper:2016cag}.
In particular, he showed how alignment without decoupling can be
achieved. George Hou presented models with  extra Higgs bosons, where alignment emerges naturally,
which should lead to distinctive signatures like triple top production~\cite{Hou:2018wch,Kohda:2017fkn}.
Margherita Ghezzi talked about doubly charged scalars, which can arise
in many BSM models, and pointed out the importance of taking finite
width effects into account~\cite{Ghezzi}.
Luc Darm{\'e} gave convincing arguments that scenarios with light  thermal dark
matter might be naturally accompanied by a corresponding 
light dark sector, offering prospects to detect a dark Higgs boson in the light spectrum~\cite{Darme:2018pop,Darme:2017glc}.
Matthew McCullough offered two different solutions to the hierarchy
problem~\cite{McCullough}, a linear dilaton model~\cite{Giudice:2017fmj} which would lead to oscillatory
signals in the $m_{\gamma\gamma}$ spectrum, and a ``hyperbolic Higgs''
model~\cite{Cohen:2018mgv}, where the Higgs boson becomes partially its own top
partner.

How to compare a plethora of BSM models to data?
{\sc Gambit}~\cite{Scott:2018jui,gambithepforge} can help: it provides a
general global fitting framework, including many statistical and
scanning options, a fast likelihood calculator, and an extensive model
database, which can be extended straightforwardly to additional models.

\section{Conclusions}

The Moriond QCD 2018 edition contained a discussion session with
the title ``Where is New Physics?'', and one named ``Heavy Flavour
indirect search for New Physics''. Should we combine this into the
slogan ``Heavy Flavour: here goes New Physics''? Even though the
flavour anomalies seem to be intriguingly persistent and consistent with
each other, it is certainly too
early make a definite statement, but the good news is that we will have
more information from the experimental side in the not too distant
future, awaiting eagerly results from LHCb and Belle II.

In any case we have seen plenty of progress to improve the precision
of Standard Model predictions in various aspects. Some results are showpieces of perturbation
theory in QCD, others operate at the interface between QCD and electroweak
corrections, or concern the deeper understanding of non-perturbative QCD.
There is also much progress in providing convenient
observables as well as tools to confront theory
predictions with data and to identify interesting regions in the vast
BSM parameter space. Some intriguing ideas about possible
extensions of the Standard Model have been presented, 
and it is not unlikely that the model builders may soon get more hints about which
direction to take.

\section*{Acknowledgments}

I am grateful to  Andrzej Czarnecki for the kind invitation, 
and I would like to thank all the organisers  of the ``Rencontres de Moriond''
for creating such a pleasant and fruitful environment. Thanks also to
the participants for discussions about their talks (and for skiing company).

\section*{References}

\bibliography{refs_moriond2018}

\end{document}